\documentclass[journal,letterpaper,twoside,twocolumn]{IEEEtran}
\usepackage{amsmath,amssymb}
\usepackage{graphicx}
\usepackage{cite}
\usepackage[usenames,dvipsnames]{color} %
\usepackage{psfrag}
\usepackage{balance}

\newcommand{\new}[1]{{#1}}

\newcommand{\C}{\mathbb C}
\newcommand{\E}{\mathbb E}
\newcommand{\pase}{P_\text{ase}}
\newcommand{\X}{\mathcal{X}}

\newcommand{\eqlab}[2]{\begin{align} \label{#1} #2 \end{align}}\newcommand{\eq}[1]{\begin{align*} #1 \end{align*}}

\newtheorem{theorem}{Theorem}
\newtheorem{lemma}[theorem]{Lemma}

\title{Influence of Behavioral Models\\on Multiuser Channel Capacity}
\author{Erik Agrell, \IEEEmembership{Senior Member, IEEE,} and Magnus Karlsson, \IEEEmembership{Fellow, OSA; Senior Member, IEEE}
\thanks{This work was supported in part by the Swedish Research Council (VR) under grants 2012-5280 and 2013-5271. The material in this paper was presented in part at the Optical Fiber Communication Conference (OFC), Anaheim, CA, Mar.~2013.

E.~Agrell is with the Dept.~of Signals and Systems, Chalmers Univ.~of Technology, SE-41296 G\"oteborg, Sweden, email agrell@chalmers.se. M.~Karlsson is with the Dept.~of Microtechnology and Nanoscience, Chalmers Univ.~of Technology, SE-41296 G\"oteborg, Sweden.}}

\begin{document}
\maketitle

\begin{abstract}
In order to characterize the channel capacity of a wavelength channel in a wavelength-division multiplexed (WDM) system, statistical models are needed for the transmitted signals on the other wavelengths. For example, one could assume that the transmitters for all wavelengths are configured independently of each other, that they use the same signal power, or that they use the same modulation format. In this paper, it is shown that these so-called behavioral models have a profound impact on the single-wavelength \new{achievable information rate}. This is demonstrated by establishing, for the first time, upper and lower bounds on the \new{maximum achievable rate} under various behavioral models, for a rudimentary WDM channel model.
\end{abstract}


\section{Introduction}
One of Shannon's most significant contributions was the definition of the \emph{channel capacity} as the highest achievable throughput (in bit/symbol or bit/s/Hz) of a given communication channel, at an arbitrarily low error probability \cite{shannon48}. He furthermore showed that a capacity-achieving transmission scheme can operate by transmitting discrete-time symbols generated from a suitably chosen \emph{input distribution,} if certain conditions are imposed on the allowed sequences of symbols.

In a more practical setting, the symbols correspond to pulses, the input distribution to a modulation format, and the allowed sequences of symbols to an error-correcting code. The maximum throughput that can be achieved with the best possible error-correcting code is, for a given channel and a given input distribution, given by the \emph{mutual information}
\cite[Ch.~2, 7]{cover06}. This quantity can be expressed as a (possibly complicated but still explicit) integral over the joint distribution of the transmitted and received symbols. Thus, it is a function of the channel and the input distribution. To obtain the channel capacity, which is a function of the channel alone, the mutual information should therefore be maximized over all possible input distributions (or modulation formats). Neither this maximization nor the mutual information integral admit analytical solutions in general,
and the exact channel capacity is therefore known only for a few specific channels, of which the additive white Gaussian noise (AWGN) channel is the most well known. This implies that for most practical channels, the capacity is only known in terms of upper and lower bounds.

For the coherent fiber-optic channel, the AWGN channel model is a good starting point, due to the amplified spontaneous emission (ASE) noise in optical amplifiers, but the nonlinearities of the optical fiber will make this channel model inaccurate for sufficiently high signal powers. Assuming the added ASE noise variance $\pase$ to be fixed and known, the question is, how will the channel capacity $C(P)$ behave as a function of the signal power $P$?
There is a common and reasonable belief \cite{splett93, mitra01,essiambre10} that the nonlinearity will somehow limit the available capacity for fiber links, but the question is to what extent.

For the single-wavelength channel, the capacity was pioneered in \cite{splett93}, where it was shown to reach a maximum and then decay as the signal power increases, and more recently referred to as the ``nonlinear Shannon limit'' \cite{ellis10, mecozzi12}. However, more or less all such plots formally represent \emph{lower bounds} on the channel capacity, as pointed out, e.g., in \cite{mitra01,turitsyn03,essiambre08,djordjevic11}, since they are obtained from analysis over a finite set over all possible input distributions \new{or using suboptimal (mismatched) receivers}. It is possible to show that the channel capacity will not decay at high signal powers, provided \new{that} a sufficiently exhaustive search \new{over} input distributions is carried out at each signal power level \cite{agrkar12,agrell12arxiv}.
Moreover, it can be shown that the use of a \emph{finite-memory channel model} will also raise the lower capacity bounds at high signal powers to nonzero values \cite{agrell14}.

In this paper, which is an extension of \cite{agrellofc13}, we will deal with the capacity of \emph{multichannel systems}, e.g., wavelength-division multiplexed (WDM) optical links, for which the situation is more subtle.
The current paradigm in optical multiuser communications \cite{mitra01,stark01,tang02,ho02,green02,kahn04,wegener04,tang06,essiambre08,freckmann09,essiambre10,ellis10,ellis11,bosco11,mecozzi12,essiambre12,secondini13} is to analyze the capacity of a single user in the system, say user 1, assuming that the other users are outside our control. We will therefore call user 1 the \emph{primary user} and the other users, whose transmissions cause interference to user 1, \emph{interferers.}
More formally, the quantity of interest is the \new{achievable information rate} $C_1 = \sup I(X_1;Y_1)$, where $I(X_1;Y_1)$ denotes the mutual information between the input $X_1$ and output $Y_1$ of subchannel $1$, and the maximization is over all possible input distributions (modulation formats) $f_{X_1}$. These quantities will be mathematically defined in Sec.~\ref{sec:infotheory}\new{, where it is also remarked that $C_1$ is in general not a channel capacity in the information-theoretic sense}. It is instructive to contrast with wireless multiuser systems, where the transmitters are typically designed jointly (but possibly operated separately), and the relevant capacity measure is a multidimensional object, the \emph{capacity region,} which describes the set of achievable throughputs for all users simultaneously \cite{taghavi06}, \cite[Ch.~15]{cover06}, \cite[Ch.~6]{elgamal11}.

\begin{figure}
\begin{center}
\includegraphics[width=\columnwidth]{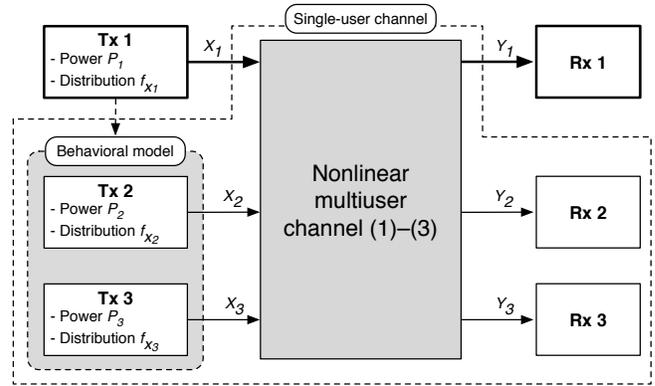}
\caption{A single-user channel model can be seen as a combination of a multiuser channel model and a behavioral model for all users but one. Transmitter and receiver are marked Tx and Rx, respectively. }\label{blockdiagram}
\end{center}
\end{figure}

Two kinds of models are needed to fully describe a multiuser system as a single-user channel model $X_1 \rightarrow Y_1$, as illustrated in Fig.~\ref{blockdiagram}: the first is a discrete-time multiuser channel model, which gives the statistics of the channel outputs $Y_1,\ldots,Y_M$ as functions of the inputs $X_1,\ldots,X_M$, and the second is a \emph{behavioral model,} which relates the interferers' distributions $f_{X_2},\ldots,f_{X_M}$ to the primary user input distribution $f_{X_1}$. Obviously, $f_{X_1}$ needs to be optimized for the considered multiuser channel model in order to attain the channel capacity, but how shall the interferers, which cause interference to the primary user, behave during this optimization process? Will they be passive, or are they allowed to adapt their signaling power and/or modulation format to the power and/or modulation format of the primary user? These questions are usually not explicity adressed in the majority of papers on optical multiuser capacity. The notable exception is the work by Taghavi \emph{et al.} \cite{taghavi06}, where both the capacity region and some bounds thereon were defined for a WDM system model, based on a Volterra approach. Their main conclusion (based on simulations of a simplified, nonlinear channel model) was that $C_1(P)$ is unbounded if the receiver could use multiuser detection to cancel nonlinear interference, and saturated (monotonically) to a constant value in the special case of increasing all user powers $P$ simultaneously.

In this paper, we discuss and classify the different behavioral models used in the literature,
and give an illustrative example of multiuser capacity for a simple nonlinear optical channel model, together with some general conclusions on how the selected behavioral model for the interferers influences $C_1(P)$. Although the idealized channel model is not fully realistic, it serves the purpose of exemplifying, for the first time, the profound impact of behavioral models on the nonlinear channel capacity. The paper is organized as follows. In Sec.~\ref{sec:system}, the multiuser nonlinear channel model is described and its parameters are defined. The behavioral models are defined in Sec.~\ref{sec:behavioral}, where we also attempt to classify the behavioral models considered in earlier optical channel capacity studies. After mathematically defining the channel capacity and related quantities in Sec.~\ref{sec:infotheory}, upper and lower bounds are derived in Sec.~\ref{sec:upper} and \ref{sec:lower}, resp. The obtained bounds are plotted and discussed in Sec.~\ref{sec:results}. The paper concludes in Sec.~\ref{sec:conclusions} with a discussion about the validity of the results and their potential extensions to more realistic optical channel models.

We use uppercase notation $X$ for random variables and lowercase $x$ for deterministic variables. Probability density functions are denoted as $f_X(x)$ and conditional probability density functions as $f_{Y|X}(y|x)$, where the subscripts will sometimes be omitted if they are clear from the context.

\section{System model} \label{sec:system}

In order to exemplify the information-theoretic nature of various behavioral models in optical communications, we need a simple, yet nontrivial, channel model for a WDM link, which enables analytical and numerical calculations of upper and lower bounds \new{on the achievable rates}. \new{Linear modulation is used in the transmitter, and the receiver applies coherent matched filtering and sampling.} We select a simplified model with three equispaced WDM channels \new{enumerated by $ i=1,2,3$.} 
\new{For simplicity, we assume that four-wave mixing dominates over self- and cross-phase modulation. This scenario arises, e.g., when the generalized phase-matching condition is fulfilled \cite{cappelini91}.}
\new{The dispersion and the nonlinearity are both assumed weak, which means that the nonlinear phase shift $\phi_\mathrm{NL} \ll 1$. Under these assumptions, the coupled nonlinear differential equations can be linearized in propagation distance by a perturbative analysis. 
A detailed discussion and the full set of coupled equations for this situation can be found in \cite{cappelini91}. We find that} the complex discrete-time output signals $Y_i$ are given by a nonlinear channel model according to
\begin{align}
Y_1 &=X_1+ \epsilon X_2^2 X_3^* +N_1, \label{sysmod1}\\
Y_2 &= X_2+ 2 \epsilon X_1 X_2^* X_3+N_2, \label{sysmod2}\\
Y_3 &= X_3+ \epsilon X_1^* X_2^2+N_3, \label{sysmod3}
\end{align}
where $X_i$ are independent, complex channel inputs and $N_i$ are independent, complex, circularly symmetric, white Gaussian noise signals, each with zero mean and equal variance. \new{The indices in \eqref{sysmod1}--\eqref{sysmod3} are the same as in \cite[Eq.~(8)]{taghavi06}, \cite[Eq.~(6)]{agrell15itw}, confining the WDM system to 3 wavelengths and ignoring self- and cross-phase modulation terms.}
\new{Similar} models were derived in the context of noncoherent WDM systems with on--off keying modulation \cite{eiselt99,forghieri97}. As in \cite{taghavi06} \new{and other works}, our intention is \emph{not} to present an accurate channel model, but rather the opposite: We wish to use the simplest possible nonlinear WDM model that will allow us to qualitatively compare different behavioral models.

In this work, we consider the \emph{single-wavelength detection} scenario, as it was defined in \cite{taghavi06}. This means that each receiver $i$ receives its own signal $Y_i$, with no information about the other received signals $Y_j$ for $j \ne i$. Furthermore, receiver $i$ knows the distributions $f_{X_j}$ of the other users $j \ne i$, but not their codebooks. Hence, \new{multiuser detection \cite{xu05,xu06,taghavi06} is possible, but not simultaneous decoding \cite[Ch.~6]{elgamal11}}.

The channel model \eqref{sysmod1}--\eqref{sysmod3} is characterized by two parameters, \new{$\epsilon$ and $\pase = \E[|N_i|^2]$}.
In an $n$-span amplified link, the single-polarization noise variance (power) equals $ \pase=n n_\text{sp} (G-1) h\nu B$, where $n_\text{sp}$ is the spontaneous emission factor, $h \nu$ the photon energy, $G$ the gain of each amplifier, which also equals the span loss, and $B$ the signal bandwidth. The constant in (\ref{sysmod1})--(\ref{sysmod3}) is $\epsilon=n \gamma L_\text{eff}$, where $\gamma$ is the fiber nonlinear coefficient and $L_\text{eff}$ the effective nonlinear amplifier span length, related to the physical amplifier separation $L$ via $L_\text{eff}=(1-\exp(-\alpha L))/\alpha$ with $\alpha$ being the fiber attenuation coefficient.
\new{One may improve the model} by multiplying $\epsilon$ with a complex factor depending on the phase mismatch, attenuation factor, and span length, but we neglect this for simplicity. 

For the numerical examples in Sec.~\ref{sec:results}, the following parameters will be used. We consider a link with $n=16$ amplifier spans. The gain of each is $G=30$ dB and the spontaneous emission factor is $n_\text{sp}=2$. The signal bandwidth is $B=40$ GHz, and with $h \nu=0.128$ aJ, $\gamma = 1.6$ W$^{-1}$ km$^{-1}$, and $L_\text{eff}=24$ km, we get $\pase=0.16$ mW and $\epsilon=610$ W$^{-1}$. \new{The condition $\phi_\mathit{NL}=\epsilon P \ll 1$, where $P$ is the signal power, translates to $P\ll1.6$ mW, or a signal-to-noise ratio of $P/\pase<10$ dB. We will apply this model, which was derived under a weak nonlinearity assumption, also in the strongly nonlinear regime, which although inaccurate is the conventional approach in the literature.}

\section{Behavioral models in multiuser communications} \label{sec:behavioral}

Whenever a multiuser system is characterized by means of a single-user channel capacity, the results are connected to a certain behavioral model, as discussed above. The behavioral models relate the input distributions of the interferers to the primary input distribution. We study three fundamentally different classes of behavioral models:
\begin{itemize}
\item[(a)] \emph{Fixed interferer distributions.} The interferer distributions $f_{X_2},\ldots, f_{X_M}$ remain the same regardless of $f_{X_1}$. The dashed arrow from Tx 1 in Fig.~\ref{blockdiagram} does not exist in this case. From the viewpoint of information theory, this is a single-user channel.
\item[(b)] \emph{Adaptive interferer power.} All users transmit with the same power $P_1=P_2=P_3$, but not necessarily the same distributions. The interferer distributions $f_{X_2},\ldots, f_{X_M}$ are fixed apart from a scale factor, which depends on $P_1$.
\item[(c)] \emph{Adaptive interferer distribution.} All users transmit with the same distribution and the same power, $f_{X_1}=f_{X_2}=f_{X_3}$.
\end{itemize}

The channel models used for WDM capacity analyses in the literature fall in categories (b) and (c). Model (b) was used by Wegener \emph{et al.} \cite{wegener04}, where on--off keying modulation was assumed for the interferers \cite[Eq.~(15)]{wegener04}, and a Gaussian pdf assumed for the primary user, although all users had the same power. \new{Behavioral model (b) was also considered in \cite[Fig.~2(b)]{secondini13}, where the influence of interferer distributions on the achievable rate of the primary user was studied. It was concluded that Gaussian interferers caused worse interference than quadrature phase-shift keying (QPSK) and ring-shaped modulation, when the primary user applies Gaussian modulation at the same power level as the interferers.}
Model (c) was used in \cite{essiambre08,essiambre10,freckmann09,taghavi06,mecozzi12,essiambre12, secondini13}, where it was explicitly stated that every channel had the same modulation and power. Multilevel ring-shaped modulation was used in \cite{essiambre08,essiambre10,freckmann09,mecozzi12,essiambre12}, \new{four different modulation formats were used in \cite[Fig.~2(a)]{secondini13},} and Gaussian modulation for all channels was used \new{in} \cite{taghavi06}.
Quite a few studies have used models of the nonlinear interference that does not depend on the choice of modulation format, but only on the power spectral density of the interferers. Then the modulation of the interferers has not been specified, and the chosen behavioral model can be either (b) or (c). This applies to \cite{mitra01,stark01,tang02,ho02,green02,kahn04,tang06,ellis10,ellis11,bosco11}.

As will be demonstrated in the following, the \new{achievable rates} may vary significantly between behavioral models.

\section{Information theory}\label{sec:infotheory}

The mutual information between two random variables $X$ and $Y$ with joint distribution $f_{X,Y}$ and marginal distributions $f_X(x) = \int f_{X,Y}(x,y) dy$ and $f_Y(y) = \int f_{X,Y}(x,y) dx$ is defined as \cite[Eq.~(2.35)]{cover06}
\eqlab{mi}{
I(X;Y) = \iint f_{X,Y}(x,y) \log_2 \frac{f_{X,Y}(x,y)}{f_X(x) f_Y(y)} dx dy
,}
where the integral is over the domain of $X$ and $Y$. If one or both of $X$ and $Y$ are discrete, their distributions are replaced with probability mass functions and the corresponding integrals are replaced with sums. Similarly, the \emph{conditional mutual information} between $X$ and $Y$ given another random variable $Z$ is defined as \cite[Eq.~(2.61)]{cover06}
\eq{
&I(X;Y|Z) \\
&\;\;= \iiint f_{X,Y|Z}(x,y|z) \log_2 \frac{f_{X,Y|Z}(x,y|z)}{f_{X|Z}(x|z) f_{Y|Z}(y|z))} dx dy dz
.}
If $X$ and $Y$ are the input and output, resp., of a communication channel, then the joint distribution can be separated into the product $f_{X,Y}(x,y) = f_X(x) f_{Y|X}(y|x)$, where $f_X$ denotes the \emph{input distribution} and $f_{Y|X}$ denotes the \emph{channel}. Thus, the mutual information depends on both the input distribution and the channel. More precisely, the mutual information gives the highest achievable rate, in bit/symbol, of a given channel and a given input distribution, if strong coding is allowed over long blocks of symbols. As discussed in the Introduction, the channel capacity is
\eqlab{infoc}{
C = \sup_{f_X} I(X;Y)
,}
which is a function of the channel only, not of the input distribution. From a practical viewpoint, the optimization over input distributions in \eqref{infoc} can be regarded as an optimization over modulation formats.

In the multiuser scenario considered in this paper, we are interested in the channel capacity of one subchannel. \new{Inspired by \eqref{infoc}, one can define}
\eqlab{infocp}{
C_i(P_i) = \sup_{f_{X_i}\!:\,\E[|X_i|^2]=P_i} I(X_i;Y_i)
,}
\new{where} $P_i = \E[|X_i|^2]=\int |x|^2 f_{X_i}(x) dx$ \new{is the signal power of subchannel $i$. This quantity is an achievable rate of subchannel $i$ and has been studied in numerous publications in optical communications. It is often called the channel capacity, although, strictly speaking, the single-user channels in Fig.~\ref{blockdiagram} have no channel capacity in an information-theoretic sense, since Shannon's channel coding theorem, according to which \eqref{infoc} gives the maximum achievable rate of the channel described by $f_{Y|X}$, assumes the channel law $f_{Y|X}$ to remain the same throughout the maximization. This is not the case in \eqref{infocp}, where $I(X_i;Y_i)$ relies on a channel law $f_{Y_i|X_i}$ that changes with $f_{X_i}$ and/or $P_i$, according to}
the behavioral models that control the input distributions \new{$f_{X_j}$} for $j\ne i$.

In this paper, we wish to evaluate \new{$C_1(P_1)$} for the behavioral models in Sec.~\ref{sec:behavioral}.%
\footnote{\new{A similar analysis can be carried out for subhannels $2$ and $3$. By symmetry, $C_3(P_3)$ is equivalent to $C_1(P_1)$, whereas $C_2(P_2)$ is different. Some of the bounds in Sec.~\ref{sec:upper} and \ref{sec:lower} extend straightforwardly to $C_2$ as well (e.g., Theorems \ref{th:mi-discrete} and \ref{th:mi-gauss}), whereas other bounding techniques, tailored to \eqref{sysmod2}, would be needed for a full characterization of $C_2(P_2)$.}}
As usual in nonlinear information theory, it seems infeasible to find an exact expression, but we can follow the standard approach and sandwich the achievable rates between upper and lower bounds. No approximations are involved in the derivations of these bounds.

\section{Upper bounds} \label{sec:upper}

Our upper bounds \new{on $C_1$} depend on the following fundamental lemma.

\begin{lemma} \label{lem:E16}
If $X$ and $Z$ are independent, then
\eq{
I(X;Y) \le I(X;Y | Z)
}
\end{lemma}

\begin{IEEEproof}
From \cite[Eq.~(2.119--120)]{cover06},
\eqlab{E16a}{
I(X;Y|Z) &= I(X;Y) + I(X;Z|Y)-I(X;Z) \\
  &= I(X;Y) + I(X;Z|Y) \label{E16b}\\
  &\ge I(X;Y) \notag
,}
where \eqref{E16a} follows from the independence of $X$ and $Z$ and \eqref{E16b} from the nonnegativity of conditional mutual information \cite[Eq.~(2.92)]{cover06}.
\end{IEEEproof}

If $X$ and $Z$ are not independent, the Lemma does not hold. A notable example is when $X \rightarrow Y \rightarrow Z$ forms a Markov chain, in which case $I(X;Y|Z) \le I(X;Y)$ follows by the data-processing inequality \cite[Eq.~(2.122)]{cover06}.

For the specific channel model \eqref{sysmod1}, the lemma can be used to derive two upper bounds.

\begin{theorem}\label{th:E21}
For any distributions of $X_2$ and $X_3$, \new{$C_1$} is upperbounded as
\eq{
C_1(P_1) \le \log_2 \left(1+\frac{P_1}{\pase} \right)
}
\end{theorem}

\begin{IEEEproof}
From \eqref{infocp} and Lemma~\ref{lem:E16},
\eqlab{E20}{
C_1(P_1) \le \sup_{f_{X_1}\!:\,\E[|X_1|^2]=P_1} I(X_1;Y_1 | X_2, X_3)
.}
Given $X_2=x_2$ and $X_3=x_3$, \eqref{sysmod1} is an AWGN channel with a constant offset $\epsilon x_2^2 x_3^*$. If this offset is known, it can be subtracted at the receiver, resulting in a regular zero-mean AWGN channel with noise variance $\E[|N_1|^2] = \pase$. Hence, the right-hand side of \eqref{E20} equals the AWGN channel capacity $\log_2(1+P_1/\pase)$, independently of $x_2$ and $x_3$, which completes the proof.
\end{IEEEproof}

Alternatively, the theorem can be derived from \eqref{sysmod1} via the data-processing inequality \cite[Th.~2.8.1]{cover06}.

Theorem~\ref{th:E21} holds for any distributions of $X_2$ and $X_3$, and therefore for any behavioral models. For certain behavioral models, the bound can be tightened using the next theorem.

\begin{theorem}\label{th:E27}
If $X_2$ and $X_3$ are zero-mean, circularly symmetric Gaussian (ZCG), then
\eq{
C_1(P_1) \le \frac{1}{P_2} \int_0^\infty e^{-u/P_2} \log_2 \left(1+\frac{P_1}{\pase+\epsilon^2 P_3 u^2} \right) du
}
\end{theorem}

\begin{IEEEproof}
Invoking Lemma~\ref{lem:E16}, this time conditioning on $X_2$ only, yields
\eqlab{E24}{
C_1(P_1) &\le \sup I(X_1;Y_1 | X_2) \notag\\
  &= \sup \int_\C f(x_2) I(X_1;Y_1 | X_2 = x_2) dx_2 \notag\\
  &\le \int_\C f(x_2) \sup I(X_1;Y_1 | X_2 = x_2) dx_2
,}
where the suprema are over all $f_{X_1}$ such that $\E[|X_1|^2]=P_1$. If $X_3$ is Gaussian, then \eqref{sysmod1} conditioned on $X_2=x_2$ is a zero-mean AWGN channel, because its two noise contributions $\epsilon x_2^2 X_3^*$ and $N_1$ are both Gaussian. The power of $\epsilon x_2^2 X_3^*$ is $\epsilon^2 |x_2|^4 P_3$, while the power of $N_1$ is $\pase$ as before. Hence, the supremum in \eqref{E24} equals the capacity of an AWGN channel with power $\pase + \epsilon^2 |x_2|^4 P_3$,
\eqlab{E25}{
C_1(P_1) &\le \int_\C f(x_2) \log_2 \left(1+\frac{P_1}{\pase+\epsilon^2 |x_2|^4 P_3} \right) dx_2
.}
This bound can be simplified by using the circular symmetry of
\eq{
f(x_2) = \frac{1}{\pi P_2} e^{-|x_2|^2/P_2}
.}
Let $U=|X_2|^2$. Then $U$ is exponentially distributed,
\eqlab{E26}{
f(u) = \frac{1}{P_2} e^{-u/P_2},\qquad u\ge 0
.}
The theorem now follows by changing the integration variable in \eqref{E25} from $x_2$ to $u=|x|^2$.
\end{IEEEproof}

\section{Lower bounds} \label{sec:lower}

Since the channel capacity is the supremum of mutual information, a lower bound on capacity can be obtained from the mutual information for any given input distribution.
\new{Analogously, from \eqref{infocp},}
\eqlab{clower}{
\new{C_1(P_1) \ge I(X_1;Y_1)}
}
for any input distribution \new{$f_{X_1}$} with power \new{$P_1$.} In this section, we will obtain lower bounds on $C_1(P_1)$ \new{via \eqref{clower}}.

If all input distributions are discrete, it is feasible to calculate the right-hand side of \eqref{clower} by numerical integration, using either of the following two theorems.

\begin{theorem}\label{th:mi-discrete}
If $X_1$, $X_2$, and $X_3$ are all discrete, uniformly distributed over complex constellations $\X_1$, $\X_2$, and $\X_3$, resp., then
\eqlab{mi-discrete}{
I(X_1;Y_1) = \E\left[ \log_2 \frac{f(y_1|x_1)}{f(y_1)} \right]
,}
where
\eqlab{fyx-discrete}{
f(y_1|x_1) &= \frac{1}{\pi \pase |\X_2| |\X_3|} \sum_{x_2\in\X_2} \sum_{x_3\in\X_3} \notag\\
&\qquad \exp\left(-\frac{|y_1-x_1-\epsilon x_2^2 x_3^*|^2}{\pase} \right), \\
f(y_1) &= \frac{1}{|\X_1|} \sum_{x_1\in\X_1} f(y_1|x_1) \label{fy-discrete}
.}
\end{theorem}

\begin{IEEEproof}
From \eqref{sysmod1},
\eqlab{fyxxx}{
f(y_1 | x_1, x_2, x_3) &= \frac{1}{\pi \pase} \exp\left(-\frac{|y_1-x_1-\epsilon x_2^2 x_3^*|^2}{\pase} \right)
.}
Marginalizing $f(y_1 | x_1, x_2, x_3)$ yields $f(y_1|x_1)$ and $f(y_1)$. Finally, \eqref{mi-discrete} follows by rewriting \eqref{mi}.
\end{IEEEproof}

\begin{theorem}\label{th:mi-gauss}
If $X_1$ is ZCG and $X_2$ and $X_3$ are discrete, uniformly distributed over complex constellations $\X_2$ and $\X_3$, resp., then
\eqlab{mi-gauss}{
I(X_1;Y_1) = \E\left[ \log_2 \frac{f(y_1|x_1)}{f(y_1)} \right]
,}
where
\eqlab{fyx-gauss}{
f(y_1|x_1) &= \frac{1}{\pi \pase |\X_2| |\X_3|} \sum_{x_2\in\X_2} \sum_{x_3\in\X_3} \notag\\
&\qquad \exp\left(-\frac{|y_1-x_1-\epsilon x_2^2 x_3^*|^2}{\pase} \right), \\
f(y_1) &= \frac{1}{\pi (P_1+\pase) |\X_2| |\X_3|} \sum_{x_2\in\X_2} \sum_{x_3\in\X_3} \notag\\
&\qquad \exp\left(-\frac{|y_1-\epsilon x_2^2 x_3^*|^2}{P_1+\pase} \right) \label{fy-gauss}
.}
\end{theorem}

\begin{IEEEproof}
In \eqref{sysmod1}, $X_1+N_1$ is ZCG with variance $P_1+\pase$, which yields
\eq{
f(y_1 | x_2, x_3) &= \frac{1}{\pi (P_1+\pase)} \exp\left(-\frac{|y_1-\epsilon x_2^2 x_3^*|^2}{P_1+\pase} \right)
.}
Marginalizing this distribution with respect to $X_2$ and $X_3$ yields $f(y_1)$ in \eqref{fy-gauss}. Equation \eqref{fyx-gauss} is proved as in the proof of Theorem~\ref{th:mi-discrete}, which completes the proof of \eqref{mi-gauss}.
\end{IEEEproof}

In Sec.~\ref{sec:results}, the expectations in \eqref{mi-discrete} and \eqref{mi-gauss} will be evaluated by Monte-Carlo integration to obtain lower bounds on \new{$C_1$} via \eqref{clower}. Theorem~\ref{th:mi-discrete} applies to all three behavioral models, as long as the interferer distributions $X_2$ and $X_3$ are discrete, whereas Theorem~\ref{th:mi-gauss} applies to some cases of models (a) and (b).

Theoretically, Theorems~\ref{th:mi-discrete} and \ref{th:mi-gauss} can be modified to hold also when at least one of the input distributions is continuous. In this case, the corresponding sums in the expressions for $f(y_1|x_1)$ and $f(y_1)$ will be replaced by integrals. However, these integrals cannot in general be evaluated analytically. This causes numerical problems in \eqref{mi-discrete} and \eqref{mi-gauss}, where the Monte-Carlo estimate of the expectation may become grossly inaccurate if $f(y_1|x_1)$ is not exact. Applying Monte-Carlo integration inside another Monte-Carlo integral should be avoided if at all possible. Therefore, we wish to find other lower bounds on the mutual information. To this end, the following lemma, due to Emre Telatar, is useful. It was stated and proved in \cite{mitra01, wegener04}, and it can also be obtained as a special case of the \emph{auxiliary-channel lower bound} \cite[Sec.~VI]{arnold06}%
\footnote{To see this, substitute $X=X_G$, $p(x) = p_G(x)$, $q(y|x) = p_G(x,y)/p_G(x)$, and $q_p(y) = p_G(y)$ in \cite[Eq.~(34)]{arnold06}.}.

\begin{lemma}\label{lem:telatar}
Let $X_G$ and $Y_G$ be complex, dependent, jointly Gaussian random variables. Let $Y$ be any complex random variable (possibly non-Gaussian) such that
\eq{
\E[|Y|^2] &= \E[|Y_G|^2], \\
\E[Y^* X_G] &= \E[Y_G^* X_G]
.}
Then
\eq{
I(X_G;Y) \ge I(X_G;Y_G)
.}
\end{lemma}

The next lemma gives the mutual information of two \new{complex,} jointly Gaussian variables. \new{It is proved by straightforward evaluation of the integral in \eqref{mi}; see, e.g., \cite[Eq.~(9-8)]{reza61}.}
\begin{lemma}\label{lem:igauss}
If $X_G$ and $Y_G$ are complex, jointly Gaussian variables with zero mean, variances $\E[|X_G|^2] = \sigma_X^2$ and $\E[|Y_G|^2] = \sigma_Y^2$, resp., and covariance $\E[X_G Y_G^*] = s_{XY}$, then their mutual information is
\eq{
I(X_G;Y_G) = \log_2 \frac{\sigma_X^2 \sigma_Y^2}{\sigma_X^2 \sigma_Y^2-|s_{XY}|^2}
.}
\end{lemma}

The preceding two lemmas make it possible to prove the following lower bound.
\begin{theorem}\label{th:E50}
For any zero-mean interferer distributions $f_{X_2}$ and $f_{X_3}$,
\eqlab{E50}{
C_1(P_1) \ge \log_2 \left( 1+\frac{P_1}{\epsilon^2 P_3 \E[|X_2|^4] + \pase} \right)
.}
\end{theorem}

\begin{IEEEproof}
Combining \eqref{clower} with Lemmas~\ref{lem:telatar} and \ref{lem:igauss} yields
\eqlab{E45}{
C_1(P_1) \ge \log_2 \frac{P_1 \sigma_Y^2}{P_1 \sigma_Y^2-|s_{XY}|^2}
,}
where
\eq{
\sigma_Y^2 &= \E[|Y_1|^2], \\
s_{XY} &= \E[X_1 Y_1^*]
,}
and $Y_1$ is given by \eqref{sysmod1} for a ZCG input distribution $X_1$. Using the independence of $X_1$, $X_2$, and $X_3$,
\eqlab{E49}{
\sigma_Y^2 &= \E[|X_1 + \epsilon X_2^2 X_3^* + N_1|^2] \notag\\
&= \E[|X_1|^2] + \epsilon^2 \E[|X_2|^4] \E[|X_3|^2] + \E[|N_1|^2] \notag\\
&= P_1 + \epsilon^2 P_3 \E[|X_2|^4] + \pase, \\
s_{XY} &= \E[X_1(X_1 + \epsilon X_2^2 X_3^* + N_1)^*] \notag\\
&= \E[|X_1|^2] \notag\\
&= P_1 \label{E48}
.}
The theorem now follows by substituting \eqref{E49}--\eqref{E48} into \eqref{E45} and simplifying.
\end{IEEEproof}

The right-hand side of \eqref{E50} depends on the statistics of $X_2$. For example, if $X_2$ is discrete, uniformly distributed over a constellation $\X_2$, then
\eqlab{E51}{
\E[|X_2|^4] = \frac{1}{|\X_2|} \sum_{x \in \X_2} |x|^4
.}
In the special case of a phase-shift keying (PSK) constellation with power $P_2$, \eqref{E51} simplifies into $\E[|X_2|^4] = P_2^2$. %

On the other hand, if $X_2$ is ZCG, then $\E[|X_2|^4]$ can be calculated by setting $X_2 = X_\text{r} + j X_\text{i}$, where $j = \sqrt{-1}$ and $X_\text{r}$ and $X_\text{i}$ are real, independent, Gaussian variables with zero mean and variance $\sigma^2 = P_2/2$. Then
\eqlab{E53}{
\E[|X_2|^4] &= \E[|X_\text{r}+j X_\text{i}|^4] \notag\\
 &= \E[X_\text{r}^4]+\E[X_\text{i}^4]+2\E[X_\text{r}^2]\E[X_\text{i}^2] \notag\\
 &= 3\sigma^4 + 3\sigma^4 + 2 \sigma^2 \sigma^2 \\
 &= 2P_2^2 \label{E54}
,}
where \eqref{E53} follows from a standard result in mathematical statistics \cite[Eq.~(5-46)]{papoulis91}.

Theorem \ref{th:E50} will be used in the next section to lower-bound \new{$C_1$} in certain cases when the interference is governed by behavioral models (a) or (b).

\section{Results} \label{sec:results}
\begin{figure}
\psfrag{C}[t][][.9]{$\new{C_1}(P_1)$}
\psfrag{SNR}[b][][.9]{$P_1/\pase$ [dB]}
\begin{center}
\includegraphics[width=\columnwidth]{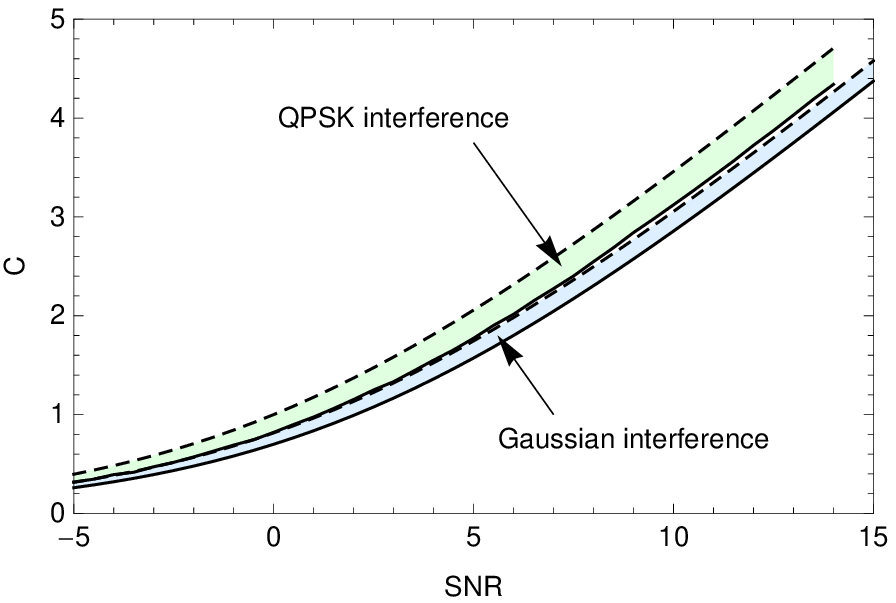} \\[-1ex]
{\footnotesize(a)} \\[2ex]
\includegraphics[width=\columnwidth]{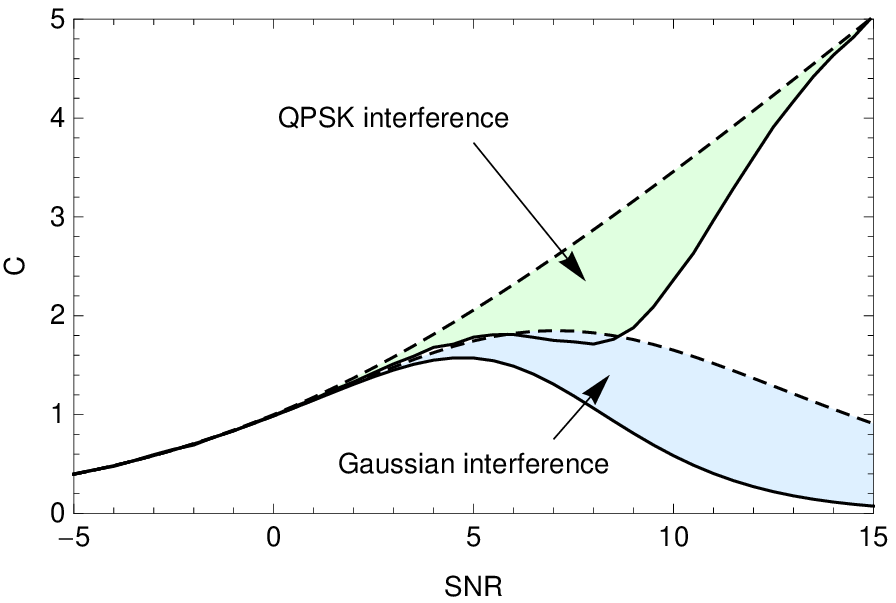} \\[-1ex]
{\footnotesize(b)} \\[2ex]
\psfrag{M2}[][][.8]{\new{$M=2$}}
\psfrag{M3}[][][.8]{\new{$M=3$}}
\includegraphics[width=\columnwidth]{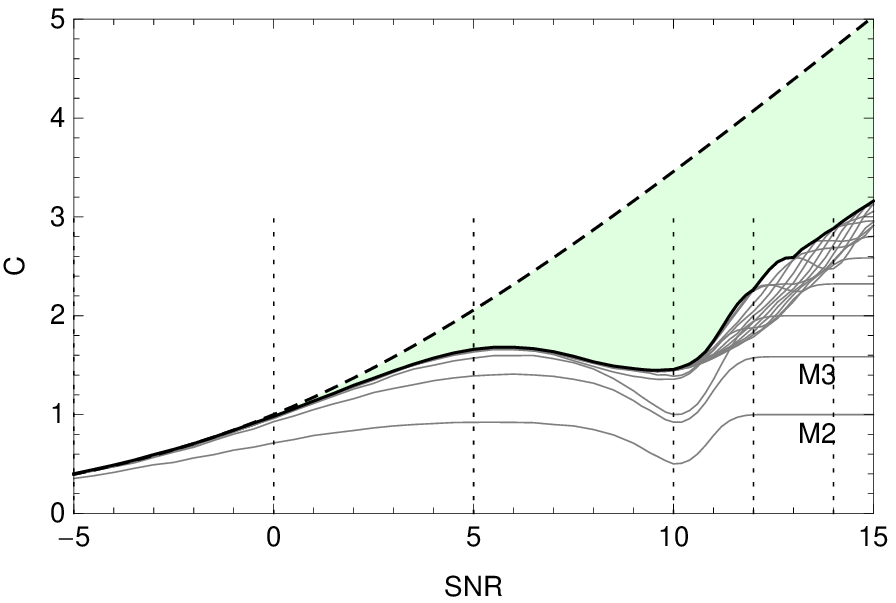} \\[-1ex]
{\footnotesize(c)}
\caption{The \new{achievable rates} $\new{C_1}(P_1)$ of user 1 in a WDM system, with the three behavioral models (a), (b), and (c), defined in Sec.~\ref{sec:behavioral}, as a function of the signal power $P_1$. Dashed lines give upper bounds and solid lines lower bounds. Shaded regions indicate the amount of uncertainty. Behavioral models (a) and (b) both have two versions, depending on the type of interferer distributions. In (c), the lower bound is obtained as the envelope of multiple bounds, indicated with gray curves. Dotted vertical lines correspond to curves in Fig.~\ref{mi-psk}.}
\label{miplots}
\end{center}
\end{figure}

In this section, the bounds of Sec.~\ref{sec:lower} and \ref{sec:upper} are numerically evaluated for the multiuser channel \eqref{sysmod1}--\eqref{sysmod3}, using the parameters $\epsilon$ and $\pase$ as specified in Sec.~\ref{sec:system}. Fig.~\ref{miplots} (a)--(c) illustrate via upper and lower bounds the single-user \new{achievable rates} $\new{C_1}(P_1) = \sup I(X_1;Y_1)$, where the maximization is over all distributions $f_{X_1}$ with power $P_1$, combined with the three behavioral models in Sec.~\ref{sec:behavioral}. For models (a) and (b), the interferer distributions $f_{X_2}$ and $f_{X_3}$ are either uniform over a \new{QPSK} constellation or Gaussian, which in total gives five scenarios. We will discuss the three models separately below.

\subsection{Behavioral model (a)---fixed interferer distributions}
With behavioral model (a), the interferer distributions $f_{X_2}$ and $f_{X_3}$ are fixed and do not change with $f_{X_1}$. The interference power is also fixed at a level of $P_2/ \pase=P_3/ \pase=5$~dB. The applied bounds are different depending on the nature of the interferers: If $X_2$ and $X_3$ follow QPSK distributions, then we obtain an upper bound from Theorem~\ref{th:E21} and a lower bound from Theorem~\ref{th:mi-gauss} or \ref{th:E50}, where Monte Carlo integration was used to estimate the expectation in \eqref{mi-gauss}. The two lower bounds turn out to be numerically indistinguishable; in Fig.~\ref{miplots} (a), Theorem~\ref{th:mi-gauss} is plotted. On the other hand, if $X_2$ and $X_3$ follow Gaussian distributions, our upper bound is given by Theorem~\ref{th:E27} and the lower bound by Theorem~\ref{th:E50} and \eqref{E54}.

The upper and lower bounds follow each other and together prove that the \new{achievable rate} increases to infinity if the signal power can be increased arbitrarily. This result is not surprising\new{, since $X_1$ dominates over the two other terms in \eqref{sysmod1} at sufficiently high power $P_1$.}
The channel is in fact a single-user channel, described by a fixed distribution $f_{Y_1|X_1}$, and the channel capacity is nondecreasing for all such channels, linear or nonlinear \cite{agrell12arxiv}. The capacity is larger in the case of discrete input distributions for the interfering channels than in the Gaussian case, but the capacity follows the same general trend in both cases.

\subsection{Behavioral model (b)---adaptive interferer power} \label{sec:adaptivepower}
With behavioral model (b), the power of all users is the same, but the distributions may be different. The same upper and lower bounds as in Fig.~\ref{miplots} (a) are plotted in Fig.~\ref{miplots} (b): Theorems~\ref{th:E21} and \ref{th:mi-gauss} with QPSK interference and Theorems~\ref{th:E27} and \ref{th:E50} with Gaussian interference. With this behavioral model,
the achievable rate of the primary channel is fundamentally different depending on the nature of the interference. If the interferers' distributions are discrete, the achievable rate increases with power towards infinity. This can be intuitively understood as follows. The magnitude of the interference term $\epsilon X_2^2 X_3^*$ will, at high enough power $P_1 = P_2 = P_3$, be much larger than $X_1$ or $N_1$. Hence, receiver 1 can detect the value of $\epsilon X_2^2 X_3^*$ with high reliability (only four values are possible in the QPSK case) and subtract this value from the received signal $Y_1$. \new{After this so-called \emph{interference cancellation,}} subchannel 1 is effectively an AWGN channel $X_1+N_1$, whose capacity is $\log_2(1+P_1/\pase)$. This is the reason why the two bounds converge near $P_1/\pase = 14$~dB and above. However, no similar receiver strategy is possible if $X_2$ and $X_3$ are Gaussian%
\footnote{As stated in Sec.~\ref{sec:system}, no receiver knows any of the other subchannels' codebooks. If these codebooks were known, the interference can be detected and substracted even for Gaussian $X_2$ and $X_3$. \cite[Sec.~15.1.5]{cover06}.}%
, because then $\epsilon X_2^2 X_3^*$, which has a continuous distribution with large variance, effectively drowns the weaker contribution from $X_1$. Therefore, this achievable rate has a peak at a moderate power, after which it decreases towards zero for very high power, as seen in Fig.~\ref{miplots} (b).

\subsection{Behavioral model (c)---adaptive interferer distribution}
To obtain a lower bound with behavioral model (c), i.e., when all users apply the same input distribution, we apply Theorem~\ref{th:mi-discrete} with a suitably chosen input distribution $f_{X_1}=f_{X_2}=f_{X_3}$. For the same reasons as in Fig.~\ref{miplots} (b), a discrete input distribution is advantageous when the interference is strong. We therefore consider $M$-PSK constellations with uniform probabilities and \new{choose the integer $M$ suitably, as described in the following.}

The bounds with behavioral model (c) are illustrated in Fig.~\ref{miplots} (c). The upper bound is again Theorem~\ref{th:E21}. The lower bound is obtained from Theorem~\ref{th:mi-discrete} as discussed in the previous paragraph. Each $M=2,\ldots,16$ gives rise to one lower bound, indicated in gray. \new{As visible in the bottom right of the figure, each of these bound converge to $\log_2 M$ at high power. This can be understood as follows. As explained in Sec.~\ref{sec:adaptivepower}, the interference term $\epsilon X_2^2 X_3^*$ in \eqref{sysmod1} can be reliably detected by reciever 1 and subtracted from $Y_1$. This holds for any discrete constellation at sufficiently high power. After interference cancellation, the effective channel is again $X_1+N_1$, whose mutual information with a uniform $M$-PSK input distribution is asymptotically $\log_2 M$. Hence, for every $M$, there exists a power threshold above which the lower bound is arbitrarily close to $\log_2 M$. This proves that the envelope of these bounds, shown in black in Fig.~\ref{miplots} (c), grows unboundedly.}

\begin{figure}
\psfrag{MI}[t][][.9]{$I(X_1;Y_1)$}
\psfrag{M}[b][][.9]{$M$}
\begin{center}
\includegraphics[width=\columnwidth]{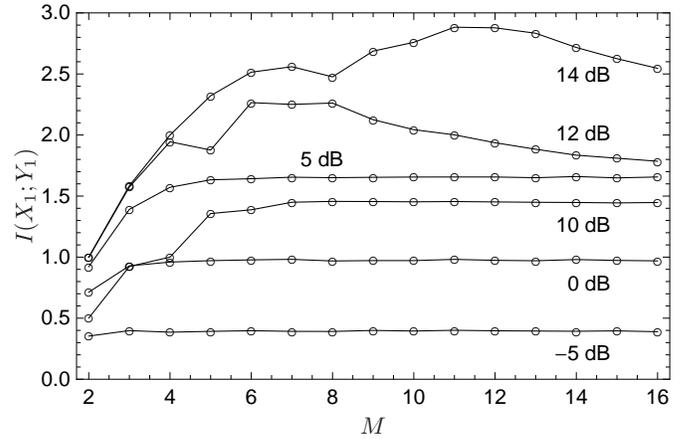}
\caption{The mutual information according to Theorem~\ref{th:mi-discrete} for $M$-PSK constellations with uniform probabilities, for the indicated values of $P_1/\pase$. The peak of each curve yields the lower bound in Fig.~\ref{miplots} (c).}
\label{mi-psk}
\end{center}
\end{figure}

\new{The} optimization process is illustrated in Fig.~\ref{mi-psk}, which shows the mutual information $I(X_1;Y_1)$ according to Theorem~\ref{th:mi-discrete} as a function of $M=2,\ldots,16$, for selected values of $P_1/\pase$. At low signal power, the mutual information is practically the same for any $M$-PSK constellation (and actually for any zero-mean distribution, including Gaussian), whereas the optimal $M$ tends to increase with power in the nonlinear regime.
We know for sure that $M$-PSK are not optimal constellations\footnote{E.g., a satellite constellation \cite{agrkar12} would improve the lower bound, at least in the range between 6 and 11~dB.}, but they suffice to show the qualitative trend of the \new{achievable rate}: It again grows with increasing power towards infinity. \new{This result is significantly stronger than} the theoretical prediction for this behavioral model \new{with arbitrary channel models \cite{agrell12arxiv}, which only states that the achievable rate is nondecreasing}.

\section{Conclusions and discussion} \label{sec:conclusions}

Multiuser information theory, or network information theory, is still in its infancy. In the information theory literature, %
the most common approach is to study the multiuser capacity region, i.e., the set of achievable rates for all users simultaneously. In this work, however, we followed the most common approach in optical communications, which is to study the channel capacity of a single user in the system. More specifically, we considered the achievable rate of a single wavelength in a multiuser WDM system, assuming certain behavioral models for the transmission on the other wavelengths.

For behavioral models (a) and (c), the \new{achievable rate} is unbounded with the signal power. With model (b), however, the outcome depends crucially on the distributions on the interfering channels; the \new{achievable rate} may increase indefinitely, as with the other behavioral models, or it may decrease to zero as the signal power increases. These results were obtained by analytically deriving both upper and lower bounds, in contrast to most previous works, which have studied lower bounds alone.

On a \new{theoretical} level, the most important conclusion in this paper is that the results depend strongly on the assumed behavioral model. We emphasize that whenever a single-channel model is derived for a multiuser system, there is always an underlying behavioral model involved. However, despite their significance, behavioral models have not yet received much attention in optical communications. Our recommendation to everyone working with the capacity of such single-user channel models is to clearly state and justify the behavioral model, because it has such a profound impact on the end results.

\new{On a more practical level, the main message is that unbounded capacity growth is indeed possible, under some specific conditions: (i) The interferers use discrete constellations; (ii) the channel model depends on the actual signals transmitted by all users, not on the statistical properties of signals \cite{agrell14, agrell15itw}; (iii) the symbol clocks of different users are synchronized; and (iv) the receiver applies multiuser detection \cite{xu05,xu06,taghavi06}.}

The results were computed for a dispersionless three-user WDM model \eqref{sysmod1}--\eqref{sysmod3}, derived in the weakly nonlinear regime. Despite its simplicity, this channel model serves to illustrate the fundamental differences between behavioral models. Future work may involve extending the channel model to the strongly nonlinear regime or accounting for dispersion, more users (wavelength channels), or dual polarization. \new{It is not known to which extent the conclusions above extend to such more realistic channels.}

\section*{Acknowledgment}

We wish to acknowledge inspiring discussions with colleagues within the Chalmers FORCE center.

\balance

\end{document}